\def\rfr#1{eq. (\ref{#1})}

\def\asec{$''$ cy$^{-1}$}

\def\dert#1#2{\frac{{{d}}{#1}}{{{d}}{#2}}}              

\def\asec{$''$ cy$^{-1}$}

\def\bar{\begin{eqnarray}}
\def\ear{\end{eqnarray}}
\def\bb{\bibitem}
\def\eqi{\begin{equation}}
\def\eqf{\end{equation}}
\def\eqia{\begin{eqnarray}}
\def\eqfa{\end{eqnarray}}
\def\rp#1#2{{#1\over#2}}

\def\lb#1{\label{#1}}

\def\oc2{$\mathcal{O}(c^{-2})$}

\def\bds#1{\boldsymbol{#1}}

\documentclass[11pt]{article}
\usepackage{amsmath,amsthm,amscd,amssymb}
\usepackage{latexsym}
\usepackage{graphicx,epsfig}
\usepackage{natbib}
\usepackage{ifpdf}
\ifpdf
\usepackage[%
  pdftitle={Instructions for use of the document class
    elsart},%
  pdfauthor={Simon Pepping},%
  pdfsubject={The preprint document class elsart},%
  pdfkeywords={instructions for use, elsart, document class},%
  pdfstartview=FitH,%
  bookmarks=true,%
  bookmarksopen=true,%
  breaklinks=true,%
  colorlinks=true,%
  linkcolor=red,anchorcolor=red,%
  citecolor=blue,filecolor=red,%
  menucolor=red,pagecolor=red,%
  urlcolor=cyan]{hyperref}
\else
\usepackage[%
  breaklinks=true,%
  colorlinks=true,%
  linkcolor=red,anchorcolor=red,%
  citecolor=blue,filecolor=red,%
  menucolor=red,pagecolor=red,%
  urlcolor=cyan]{hyperref}
\fi

\begin{document}

\noindent{\bf \LARGE{Astronomical constraints on some long-range  models of modified gravity}}
\\
\\
\\
{Lorenzo Iorio}\\
{\it Viale Unit$\grave{\it a}$ di Italia 68, 70125\\Bari (BA), Italy
\\tel. 0039 328 6128815
\\e-mail: lorenzo.iorio@libero.it}

\begin{abstract}
In this paper we use the corrections to  the usual Newton-Einstein secular precessions of the perihelia of the inner planets of the Solar System, phenomenologically estimated as solve-for parameters by the Russian astronomer E.V. Pitjeva by fitting almost one century of data with the EPM2004 ephemerides,
in order to constrain some long-range  models of modified gravity recently put forth to address the dark energy and dark matter problems. The models examined here are the four-dimensional ones obtained with the addition of inverse powers and logarithm of some curvature invariants, and the multidimensional braneworld model by Dvali, Gabadadze and Porrati (DGP). After working out the analytical expressions of the secular perihelion precessions  induced by the corrections to the Newtonian potential of such models, we compare them to the estimated corrections to the rates of perihelia by taking  their ratio for different pairs of planets  instead of using one perihelion at a time for each planet separately, as done so far in literature. As a result, the curvature invariants-based models are ruled out, even by re-scaling by a factor 10 the errors in the planetary orbital parameters estimated by Pitjeva. Less neat is the situation for the DGP model. Only the general relativistic Lense-Thirring effect, not included, as the other exotic models considered here, by Pitjeva in the dynamical force models used in the estimation process, passes such a test. It would be important to repeat the present analysis by using corrections to the precessions of perihelia independently estimated by other teams of astronomers as well, but, at present, such rates are not yet available.
\end{abstract}



 \section{Introduction}\label{intro}
 Weak-field limits of theories involving long-range modifications of gravity recently put forth to address the issues of dark energy and dark matter \citep{DGP,All05,NvA05,NvA06,NvA06b,Apo06,Cap06,Noj07} are important because such exotic corrections to the Newtonian potential allow, in principle, for tests to be performed on local, astronomical scales, independently of the galactic/cosmological effects \citep{Cap07} which motivated such alternative theories and that, otherwise, would represent their only justification.  In this paper we will show how to obtain phenomenologically tight constraints on the viability of some of such modified theories by suitably using the latest observational results from Solar System planetary motions \citep{Pit05a, Pit05b}.

 In Section \ref{alf} we will consider the  four-dimensional models obtained from inverse powers of some curvature invariants \citep{NvA05}. After working out the analytic expression of the secular, i.e. averaged over one orbital revolution, perihelion precession induced by the Newtonian limit of such models, we will compare it to the phenomenologically estimated corrections to the usual Newton-Einstein precessions of the perihelia of the inner planets of the Solar System. By taking the ratio of them for different pairs of planets we will find that the predicted exotic effects are ruled out.  In Section \ref{bet} and Section \ref{gam} we repeat the same procedure for the four-dimensional model based on the logarithm of some invariants of curvature \citep{NvA06b} and for the multidimensional braneworld model by \citet{DGP}, respectively, by finding that also such models do not pass our test. In Section \ref{delt} we apply the same strategy to the general relativistic gravitomagnetic field \citep{LT} finding that it is, instead, compatible with the ratio of the perihelion precessions for all the considered pairs of planets.
 Section \ref{concl} is devoted to the conclusions.

 \section{The inverse-power curvature invariants models}\lb{alf}
 In this Section we will address the long-range modifications of gravity obtained by including in the action  inverse powers of some invariants of curvature not vanishing in the Schwarzschild solution \citep{NvA05,NvA06,NvA06b}.
 From  the correction to the Newtonian potential \citep{NvA05}
\eqi V = \frac{\alpha GM}{ {r_c}^{6k+4} }r^{6k+3},\ r\ll{r_c}, \lb{pot}\eqf
where $k$ is a positive integer number,
it follows a purely radial acceleration
\eqi\bds A = -\rp{\alpha GM(6k+3)}{{r_c}^{6k+4}}r^{6k+2}\ \bds{\widehat{r}},\ r\ll{r_c} \lb{aAvA}.\eqf
The length scale $r_c$ depends, among other things, on a parameter $\mu$ which must assume a certain value in order that the model by \citet{NvA05} is able to reproduce the cosmic acceleration \citep{Car05,Mena06} without dark energy; it is just such a value of $\mu$ which makes $r_c \approx 10$ pc ($k=1$) for a Sun-like star \citep{NvA05}.
Since \citep{NvA06}\eqi\alpha =\rp{k(1+k)}{(6k+3)2^{4k}3^k}\eqf and ${r_c}\approx 10$ pc ($k=1$),  the condition $r\ll{r_c}$ for which the expansion in $r/r_c$ yielding \rfr{pot} retains its validity is fully satisfied in the Solar System, and \rfr{aAvA} can be treated as a small correction to the Newtonian monopole with the standard perturbative techniques of celestial mechanics.
The Gauss equation for the variation of the pericentre $\omega$ of a test particle acted upon by an entirely radial disturbing acceleration $A$ is
\eqi\dert\omega t = -\rp{\sqrt{1-e^2}}{nae}A\cos f,\lb{gaus}\eqf
where $a$ and $e$ are the semimajor axis and the eccentricity, respectively, of the orbit of the test particle, $n=\sqrt{GM/a^3}$ is the unperturbed Keplerian mean motion  and $f$ is the true anomaly reckoned from the pericentre.
The secular precession of the pericentre $\left\langle\dot\omega\right\rangle$ can be worked out by evaluating the right-hand side of \rfr{gaus} onto the unperturbed Keplerian ellipse \eqi r=a(1-e\cos E),\eqf where $E$ is the eccentric anomaly, and by performing subsequently an integration over one full orbital period
To this aim, the following relations are useful
\eqi dt = \left(\rp{1-e\cos E}{n}\right)dE,\eqf
\eqi \cos f = \rp{\cos E-e}{1-e\cos E}.\eqf

Let us start with the case $k=1$;
the extra-acceleration becomes
\eqi \bds A_{k=1}=-\rp{9\alpha GM}{{r_c}^{10}}r^8\ \bds {\widehat{r}}.\eqf
By proceeding as previously outlined and using  the exact result
\eqi\int_0^{2\pi}(\cos E-e)(1-e\cos E)^8 dE=-\rp{5e\pi}{64}\left[128+7e^2\left(128+160e^2+40e^4+e^6\right)\right],\eqf
it is possible to obtain  the exact formula
\eqi\left\langle\dot\omega\right\rangle_{k=1} = -\rp{45\alpha}{{r_c}^{10}}\sqrt{GMa^{17}(1-e^2)}\left[1+7e^2\left(1+\rp{5}{4}e^2 +\rp{5}{16}e^4 +\rp{1}{128}e^6 \right)\right]\lb{peri}.\eqf
It is important to note the dependence of $\left\langle\dot\omega\right\rangle$ on a positive power of the semimajor axis $a$: this fact will be crucial in setting our test.

The predicted extra-precession of \rfr{peri} can be fruitfully compared to the corrections to the usual Newton-Einstein perihelion rates of the inner planets of the Solar System phenomenologically estimated by \citet{Pit05a}, in a least-square sense, as solve-for parameters of a global solution in which a huge amount of modern planetary data of all types  covering about one century were contrasted to the dynamical force models of  the EPM2004 ephemerides \citep{Pit05b}.  Such corrections are quoted in Table \ref{tavola}.
\begin{table}
\caption{Semimajor axes $a$, in AU (1 AU$=1.49597870691\times 10^{11}$ m), and phenomenologically estimated corrections to the Newtonian-Einsteinian perihelion rates, in arcseconds per century (\asec), of Mercury, the Earth and Mars  \citep{Pit05a}. Also the associated errors are quoted: they are in m for $a$ \citep{Pit05b} and in \asec\ for $\dot\varpi$ \citep{Pit05a}. For the semimajor axes they are the formal, statistical ones, while for the perihelia they are realistic in the sense that they
were obtained from comparison of many different
solutions with different sets of parameters and observations (Pitjeva, private communication 2005). The results presented in the text do not change if $\delta a$ are re-scaled by a factor 10 in order to get more realistic uncertainties.}\label{tavola}

\begin{tabular}{ccccc} \noalign{\hrule height 1.5pt}
Planet & $a$ (AU) & $\delta a$ (m)  & $\dot\varpi$ (\asec) & $\delta\dot\varpi$ (\asec) \\
\hline
Mercury & 0.38709893 & 0.105 & -0.0036 & 0.0050\\
Earth & 1.00000011 & 0.146 & -0.0002 & 0.0004 \\
Mars & 1.52366231 & 0.657 & 0.0001 & 0.0005\\

\hline

\noalign{\hrule height 1.5pt}
\end{tabular}

\end{table}
 They were determined in a model-independent way, without modeling this or that particular model of modified gravity: only known Newton-Einstein accelerations\footnote{With the exception of the general relativistic gravitomagnetic interaction, yielding the Lense-Thirring effect, and of the Kuiper Belt Objects.} were, in fact, modeled so that the estimated perihelion extra-rates account, in principle, for all the unmodeled forces present in Nature.
 Since July 2005 \citep{Ior07a}, many other authors so far  used  the extra-precessions of the perihelia of the inner planets of the Solar System estimated  by \citet{Pit05a}  to put constraints on modified models  of gravity \citep{Gan06,IorDGPb,San06,Ad07,Rug07,Ior07b}, cosmological constant \citep{Ior06a,Ser06b}, various cosmological issues \citep{Adetal07,Fay07,Nes07a,Nes07b,Ser07}, dark matter distribution \citep{Ior06b,Khri06,Ser06a,Fre07,Khri07}, trans-Neptunian bodies \citep{Ior07c}, general relativity \citep{Wil06,Ior07a}; a common feature of all such analyses is that they always used the perihelia separately for each planet, or combined  linearly by assuming that the exotic effects investigated were included in the estimated corrections to the perihelia precessions, and by using their errors to constrain the parameters of the extra-forces. About the reliability of the results by \citet{Pit05a}, \citet{Ior07b} made an independent check by assessing the total mass of the Kuiper Belt Objects and getting results compatible with other ones obtained with different methods, not based on the dynamics of the inner planets.
 It must  be noted that   more robustness could be reached if and when other teams of astronomers will estimate their own corrections to the perihelion precessions. On the other hand, an alternative approach would consist in re-fitting the entire data set by including an ad-hoc parameter accounting for just the exotic effect one is interested in. However, such a procedure might be not only quite time-consuming because of the need of modifying the software's routines by including the extra-accelerations, but it would be also model-dependent by, perhaps, introducing  the temptation of more or less consciously tweaking somehow the data and/or the procedure in order to obtain just the outcome one a-priori expects.

 Here we will not use one perihelion at a time for each planet. Indeed, let us consider a pair of planets A and B and take the ratio of their estimated extra-rates of perihelia: if \rfr{peri} is responsible for them, then the quantity\footnote{It turns out that the multiplicative term depending on the eccentricities has a negligible effect on our conclusions.}  \eqi \Gamma_{\rm AB} =\left|\rp{\dot\omega^{\rm A}}{ \dot\omega^{\rm B} }- \left(\rp{a^{\rm A}}{a^{\rm B}}\right)^{17/2} \right|\lb{gam}\eqf  must be compatible with zero, within the errors.   The figures of Table \ref{tavola} tell us that it is definitely not so: indeed, for
A=Mars, B=Mercury we have
\eqi\Gamma_{\rm MaMe}=10^5\pm 0.1.\eqf The situation is slightly better for A=Mars and B=Earth:
\eqi \Gamma_{\rm MaE}=38\pm 3.5.\eqf
An intermediate case occurs for A=Earth and B=Mercury:
\eqi\Gamma_{\rm EMe}=10^3 \pm 0.2.\eqf It is important to note     that
\begin{itemize}
\item The uncertainty in $\Gamma_{\rm AB}$ has been conservatively estimated as
\eqi\delta\Gamma_{\rm AB}\leq \left|\rp{\dot\omega^{\rm A}}{\dot\omega^{\rm B}}\right|\left(\rp{\delta\dot\omega^{\rm A}}{|\dot\omega^{\rm A}|} + \rp{\delta\dot\omega^{\rm B}}{|\dot\omega^{\rm B}|}\right) + \rp{17}{2}\left(\rp{a^{\rm A}}{a^{\rm B}}\right)^{17/2}\left(
\rp{\delta a^{\rm A}}{a^{\rm A}} + \rp{\delta a^{\rm B}}{a^{\rm B}} \right)\eqf
by linearly adding the individual terms coming from the propagation of the errors in $\dot\omega$ and $a$ in \rfr{gam}; this is justified by the existing correlations among the estimated Keplerian orbital elements\footnote{The correlations among the perihelion rates are low, with a maximum of 20$\%$ between the Earth and Mercury (Pitjeva, private communication, 2005).}

    \item The results presented here do not change if we re-scale by a factor 10 the formal errors in the semimajor axes \citep{Pit05b} quoted in Table \ref{tavola}. The same holds also for the errors in the perihelia rates which, however, are not the mere statistical ones but are to be considered as realistic, as explained in the caption of Table \ref{tavola}

        \item The constraints obtained here with \rfr{gam} are independent of $\alpha$ and ${r_c}$; should one use \rfr{peri} for each planet separately to constrain ${r_c}$, it turns out that, for $\alpha=4\times 10^{-3}$ ($k=1$), ${r_c}\lesssim 4.5$ AU. Note that with such a value the condition $r\ll{r_c}$, with which \rfr{pot} and, thus, \rfr{peri} were derived, holds   for all the inner planets

        \item For $k>1$ the situation is even worse because of the resulting higher powers with which the semimajor axis  enters the formulas for the perihelion rates
\end{itemize}

\section{The logarithmic curvature invariants models}\lb{bet}
The same approach can be fruitfully used for the model by \citet{NvA06b} based on an action depending on the logarithm of some invariants of the curvature in order to obtain a modification of gravity at the MOND \citep{Mil83} characteristic scale \citep{San02}, so to address in a unified way the dark energy and dark matter problems; in this model the length scale $r_c$  amounts to about 0.04 pc for the Sun. The correction to the Newtonian potential  is
\eqi V\propto \rp{GMr^3}{{r_c}^4},\eqf which yields the perturbing acceleration
\eqi \bds A \propto \rp{r^2}{{r_c}^4}\bds {\widehat{r}}.\lb{newac}\eqf By using
\eqi \int_0^{2\pi}(\cos E-e)(1-e\cos E)^2dE = -e\pi(4+e^2),\eqf the secular precession of perihelion induced by \rfr{newac} is
\eqi\left\langle\dot\omega\right\rangle \propto \rp{\sqrt{GM a^5 (1-e^2)}}{{r_c}^4}(4+e^2);\lb{logperi}\eqf  also in this case it depends on a positive power of the semimajor axis; cfr. the approximated result by \citet{NvA06b} for the shift per orbit, i.e. $2\pi\left\langle\dot\omega\right\rangle /n$.

By taking the ratio of \rfr{logperi} for a pair of planets and comparing it to the ratio of the estimated extra-precessions by \citet{Pit05a} it can be obtained
\eqi \Delta_{\rm AB} = \left|\rp{\dot\omega^{\rm A}}{ \dot\omega^{\rm B} }- \left(\rp{a^{\rm A}}{a^{\rm B}}\right)^{5/2} \right|\lb{DEL}. \eqf
The test is not passed. Indeed, for A=Mars and B=Mercury we have
\eqi \Delta_{\rm MaMe}=30.7\pm 0.1;\eqf
the pair A=Earth, B=Mercury yields
\eqi \Delta_{\rm EMe}=10.6\pm 0.2,\eqf
while A=Mars, B=Earth $\Delta$ is marginally compatible with zero
\eqi \Delta_{\rm MaE}=3.4\pm 3.5.\eqf
Note that, even if the real errors in the estimated extra-precessions of perihelia were up to 10 times larger than those quoted by \citet{Pit05a},
the pair Mars-Mercury would still be able to rule out the logarithmic model by \citet{NvA06b}.
\section{The multidimensional braneworld Dvali-Gabadadze-Porrati model}\lb{gam}
Another modified model of gravity aimed to explain the cosmic acceleration without dark matter is the multidimensional braneworld model DGP \citep{DGP} which predicts, among other things, an extra-rate of perihelion independent of the planetary semimajor axis\footnote{The only dependence on the features of the planetary orbits occurs through a correction quadratic in the eccentricity $e$ \citep{IorDGP} which turns out to be negligible in this case.} \citep{LS,IorDGP}. It is incompatible with the test of the ratio of perihelia as well, although less dramatically than the previously examined models. Indeed,  by defining
\eqi\Psi_{\rm AB}=\left|\rp{\dot\omega^{\rm A}}{\dot\omega^{\rm B}}-1\right|,\eqf
for A=Mars, B=Mercury  we have
\eqi \Psi_{\rm MaMe}=1.0\pm 0.2,\eqf
while A=Earth, B=Mercury yield
\eqi \Psi_{\rm EMe}=0.9\pm 0.2.\eqf
Errors in the determined extra-rates of perihelion 5 times larger than those quoted in Table \ref{tavola} would allow the DGP model to pass the test.
The pair A=Mars, B=Earth   give a result compatible with zero:
\eqi \Psi_{\rm MaE}=1.5\pm 3.5;\eqf the same hold for the other three combinations in which A and B denotes the planets with the smaller and larger semimajor axes, respectively.
Until now, the DGP model was not found in disagreement with the Solar System data because the perihelia were used separately for each planet    \citep{IorDGPb}.

\section{General relativistic effects: gravitomagnetism and the cosmological constant}\lb{delt}
It maybe interesting to note that, contrary to the exotic effects  induced by the modified models of gravity previously examined, the Lense-Thirring effect \citep{LT} induced by the general relativistic gravitomagnetic field of the Sun, not modeled by \citet{Pit05a}, does pass our test based on the ratio of the perihelia.
Indeed, since the Lense-Thirring perihelion precessions are proportional to a negative power of the semimajor axis, i.e.
\eqi\left\langle\dot\omega\right\rangle\propto a^{-3},\eqf
the quantity
\eqi \Lambda_{\rm AB}=\left|\rp{\dot\omega^{\rm A}}{\dot\omega^{\rm B}}-\left(\rp{a_{\rm B}}{a_{\rm A}}\right)^3\right|\eqf must be considered. It turns out that it is compatible with zero for all the six combinations which can be constructed with the data of Table \ref{tavola}. This result enforces the analysis by \citet{Ior07a} in which the extra-rates of the perihelia were used one at a time for each planet and linearly combined by finding the general relativistic predictions for the Lense-Thirring precessions compatible with them.


\section{Conclusions}\lb{concl}
In this paper we used the corrections to the Newton-Einstein secular precessions of the perihelia of the inner planets of the Solar System,   estimated by \citet{Pit05a}    in a least-square sense as phenomenological solve-for parameters of a global solution in which almost one century of data were fitted with the EPM2004 ephemerides, to put tight constraints on several models of modified gravity recently proposed to explain dark energy/dark matter issues. By using the ratio of the perihelion precessions for different pair of planets, instead of taking one perihelion at a time for each planet as done so far, we were able to rule out all the considered long-range  models of modified gravity, in particular the ones based on inverse powers of curvature invariants by \citet{NvA05} and on the logarithm of some curvature invariants  \citep{NvA06b}, even by re-scaling by a factor 10 the errors in the estimated perihelion extra-rates. The situation is less dramatic for the DGP \citep{DGP} braneworld model  since if the real errors in the perihelion precessions were, in fact, 5 times larger than the ones released it would become compatible with the data.
Only the general relativistic Lense-Thirring effect passed the test.  However, it must be noted that our results are based only on the extra-rates of perihelia determined by \citet{Pit05a}: it would be highly desirable to use corrections to the secular motion of perihelia estimated by other teams of astronomers as well. If and when they will be available  our test will become more robust.

\section*{Acknowledgments}
I am grateful to G.E. Melki for useful remarks.


\end{document}